\documentclass[amsmath, amssymb, reprint]{revtex4-1}

\usepackage{graphicx}
\usepackage{dcolumn}
\usepackage{bm}
\usepackage{hyperref}

\begin{document}


\title{Interfacial thermodynamics of spherical nanodroplets: Molecular understanding of surface tension via hydrogen bond network}

\author{QHwan Kim}
\author{Wonho Jhe}%
 \email{whjhe@snu.ac.kr}
\affiliation{%
 Center for 0D Nanofluidics, Department of Physics and Astronomy, Institute of Applied Physics,  Seoul National University, Gwanak-gu, Seoul 08826, Republic of Korea.
}%


\date{\today}

\begin{abstract}
Surface tension plays a ubiquitous  role in  phase transitions including condensation or evaporation of atmospheric liquid droplets. Especially,  understanding of interfacial thermodynamics of the critical nucleus of 1 nm scale is important for  molecular characterization of the activation energy barrier of nucleation. Here, we investigate surface tension of spherical nanodroplets with both molecular dynamics and density functional theory, and find that surface tension decreases appreciably below 1 nm radius, whose analytic expression is consistently derived from the classic Tolman's equation. In particular, the free energy analysis of nanodroplets shows that the change of surface tension originates dominantly from the configurational energy of interfacial molecules, which is evidenced by the increasingly disrupted hydrogen bond network as the droplet size decreases. Our result can be applied to the interface-related phenomena associated with molecular fluctuations such as biomolecule adsorption at sub-nm scale where the macroscopic thermodynamic quantities are ill-defined.
\end{abstract}

\pacs{Valid PACS appear here}
\maketitle


Nucleation of nanoscale water droplets in the atmospheric clouds is crucial for the control of earth climate \cite{Baker1997, Kulmala2004}. When homogeneous nucleation initiates with the formation of water nanodroplets from the supersaturated vapor, there exists the thermodynamic energy barrier formed by the difference between the favorable volume- and unfavorable surface-energy. The barrier height, which determines the nucleation rate of the condensed phase, is governed by the surface tension of the critical nucleus \cite{Kalikmanov2005}. The classical nucleation theory has employed the value of surface tension of planar interface in estimating the nucleation rate and has suffered from poor predictability, which is even orders of magnitude off from the experimental results of  liquid droplets such as the simple liquid and water \cite{Oxtoby1988, Holten2005, Bruot2016}. Moreover, the surface energy model with the planar tension produces (i)  inaccurate estimation of the solvation free energy of hydrophobic molecules \cite{Sedlmeier2012} and (ii) improper description of the heterogeneous nucleation processes of the concave nanodroplets formed between two hydrophilic surfaces \cite{Kwon2018, Kim2018}. Despite numerous theoretical \cite{Factorovich2014, Sampayo2010,Koga1998, Malijevsky2012, Rehner2018} and experimental \cite{Bruot2016, Kim2018, Kwon2018} studies performed for decades, determination of surface tension of the curved interface \cite{Bruot2016,  Rowlinson1982, Joswiak2013, Azouzi2013, Wilhelmsen2015, Kwon2018, Kim2018-2} of the nucleus is still a topic of ongoing controversies. Therefore, quantitative and consistent study of surface tension of the highly curved liquid-vapor  interface should be carried out for the fundamental understanding of the nucleation phenomena. 

 The first attempt to describe the curvature dependence of surface tension was made by the Tolman's equation in 1949 \cite{Tolman1949},
\begin{equation}
 \label{eqn:Tolman}
 \frac{\gamma(R_{\textrm{s}})}{\gamma_{0}} = \frac{1}{1+2\delta_{0}/R_{\textrm{s}}} = 1-\frac{2\delta_{0}}{R_{\textrm{s}}}+ O\left(\frac{1}{R_{\textrm{s}}^{2}}\right),
 \end{equation}
 where the constants $\gamma_{0}$ and $\delta_{0}$ are the surface tension and Tolman length of the planar interface, respectively. Notice that $\delta_{0}$ is defined as the difference between  two possible definitions of the nanodroplet radius, equimolar radius $R_{\textrm{e}}$ and the radius of surface of tension $R_{\textrm{s}}$, when the nanodroplets become infinitely large;  $\delta_{0}  \equiv  \lim_{R_s\rightarrow\infty}\delta(R_{\textrm{s}}) = \lim_{R_s\rightarrow\infty}\left(R_{\textrm{e}} - R_{\textrm{s}}\right)$. Therefore, Tolman's equation provides a general description of  surface tension of nanodroplets in terms of only a `single' parameter $\delta(R_{\textrm{s}})$ for a given $R_{\textrm{s}}$. Theoretical studies suggest that the magnitude of the constant $\delta_{0}$  is comparable to  the molecular diameter \cite{Rowlinson1982}, but there are still many debates concerning the magnitude and sign of $\delta_{0}$. Moreover, despite substantial development of experimental methods \cite{Bruot2016,Kim2018}, there does not  exist a consensus on either the value of $\delta_{0}$ nor the change of surface tension at the molecular scale because quantitative information of  molecules in the nanodroplets still lacks, which may behave very differently from the planar case. Therefore, for a unified understanding of the molecular-scale surface tension and the nanoscale liquid-vapor interface, single-molecule study is on high demand, including the investigation of the free energy as well as hydrogen bond (HB) networks of individual molecules. 
 
 Here, we employ the molecular dynamics (MD) and density functional theory (DFT) to calculate the surface tension of liquid nanodroplet from the pressure profile and the thermodynamic free energy. We find curvature dependence of surface tension below 1 nm radius and derive analytic expression from the classic Tolman's equation combined with the radius-dependent Tolman length \cite{Tolman1949}. The derived analytic expression predicts surface tension within the error below 1$\%$. Moreover, to describe the origin of the change of surface tension in the nanoscale droplet, we use the statistical analysis of individual  molecules that provides the free energy of nanodroplets and also use the HB network analysis with graph theory approach. They show that surface tension is governed by the energy change of interfacial molecules, associated with the disruption of the HB network.

\begin{figure}
\includegraphics[width=8.6cm]{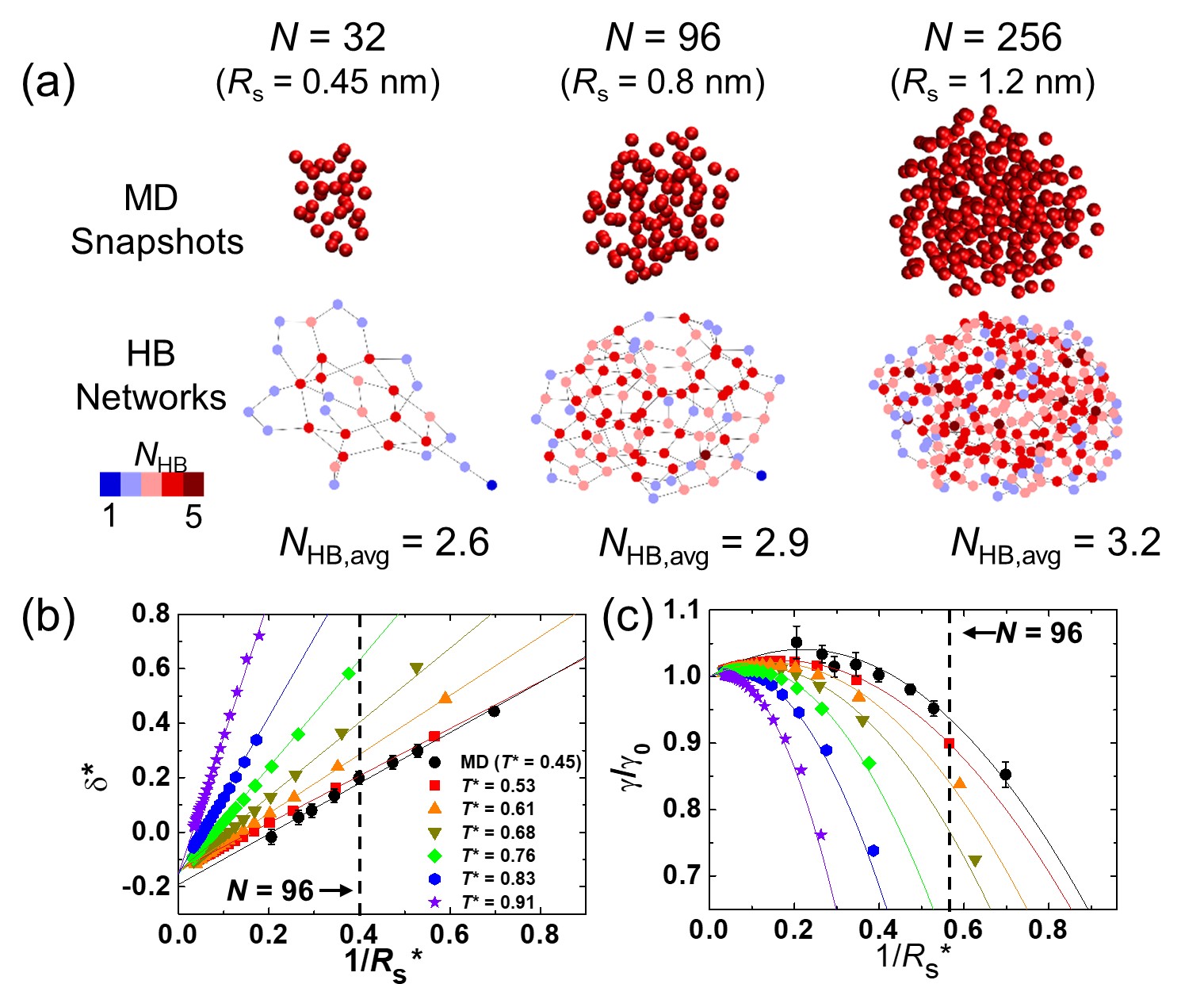}
\centering
\caption{Hydrogen bond network, Tolman length and surface tension of nanodroplets. (a) MD snapshots (top row) and visualization of hydrogen bond (HB) networks (bottom row) of $N = 32, 96, 256$ nanodroplets. The node color represents the HB number of water molecules, $N_{\textrm{HB}}$, from 1 (blue) to 5 (red). The average numbers of HB, $N_{\textrm{HB,avg}}$, are given. (b) MD and DFT simulation results (dots) of the Tolman length $\delta^{*}$ as a function of the inverse radius  $1/R_{\textrm{s}}^{*}$. The lines represent the fitting of the results with the ansatz $\delta^{*}\left(R_{\textrm{s}}^{*}\right) = \delta_{0}^{*} + {\delta_{1}^{*}}/{R_{\textrm{s}}^{*}}$. The vertical dashed line corresponds to $N= 96$ or $R_{\textrm{s}}$ = 0.8 nm.  (c) The mutual agreement of surface tension between the simulation results (dots) and Eq. \ref{eqn:TtoH} (curves) exists.}
\label{fig:fig1}
\end{figure}

Figure \ref{fig:fig1}(a) summarizes the main results of our work.
The MD snapshots of nanodroplets for $N = 32$, 96 and 256 water molecules (or equivalently $R_{\textrm{s}} = 0.45$, 0.8 and 1.2 $\textrm{nm})$  are provided in the top row. The bottom row  shows the network representations of the snapshots.
 The points (nodes) and lines (edges)  of the HB network correspond to the water molecules and HBs, respectively. The strength of the interaction is qualitatively represented by the number of HB, $N_{\textrm{HB}}$, which is reflected by the color of the nodes. The color distributions in the networks and the average values of the HB number, $N_{\textrm{HB,avg}}$, demonstrate that the overall interaction strength, which is directly related to the surface tension of nanodroplets,  decreases considerably when $R_{\textrm{s}} < 1 \ \textrm{nm}$.

To study nanodroplet systems, we have performed the MD simulation of the TIP4P/2005 water model and the DFT calculation of the Lennard-Jones liquid (refer to Supplementary Information for detailed information about MD and DFT). In the MD simulation, the number of water molecules ranges from $N = 32$ to 512 at $T$ = 290 K. In DFT, we deal with nanodroplets with a radius of $R^{*} = 1.77$ to 30 at the temperature $T^{*} = 0.53$ to  0.91. We chose the smallest radius of droplet in DFT as the lower size limit without thermal instability. In our simulations, the reduced units are employed: $r^{*} = r/\sigma$, $R^{*}=R/\sigma$, $\delta^{*} = \delta/\sigma$, $\rho^{*}=\rho\sigma^{3}$ and $T^{*}=T/T_{c}$, where $\sigma$ is the diameter of the liquid molecule and $T_{c}$ is the critical temperature of the liquid. For the TIP4P/2005 water model \cite{Vega2006}, we used  $T_{c} = 641.4$ K and $\sigma = 0.3159$ nm. 
 
Let us first calculate the theoretical values of the Tolman length, $\delta^{*}(R_{\textrm{s}}^{*}) = R_{\textrm{e}}^{*} - R_{\textrm{s}}^{*}$, which are presented in Fig. \ref{fig:fig1}(b) as a function of the inverse radius of the surface of tension, $1/R_{\textrm{s}}^{*}$ (definitions of  $R_{\textrm{e}}^{*}$  and $R_{\textrm{s}}^{*}$ are summarized in Supplementary Information). For the curvature dependence,  we have adopted (and justified later) the ansatz in the linear form, $\delta^{*}\left(R_{\textrm{s}}^{*}\right) = \delta_{0}^{*}+\delta_{1}^{*}/R_{\textrm{s}}^{*}$, which  fits well with the MD and DFT results\cite{Bruot2016, Block2010}. The  resulting numerical values of the coefficients $\delta_{0}^{*}$ and $\delta_{1}^{*}$ are listed in Table \ref{table:tolman}.   Interestingly, at $T^{*}$ = 0.45 (water droplet), the sign of the Tolman length changes from negative to positive at $R_{\textrm{s}} \sim 1.2 \ \textrm{nm}$, which marks the region where surface tension begins to decrease. Notice  that the Tolman length we find for the planar interface of water is $\delta_{0} = -0.59 \ \mathring{\textrm{A}}$, which is consistent with the previous theoretical and experimental results\cite{Wilhelmsen2015, Azouzi2013, Joswiak2013}.

  
 \begin{table}
   \centering
    \caption{Tolman length and bending rigidity of nanodroplets of MD and DFT results}
    \label{table:tolman}
     \begin{tabular}{c c c c}
          \hline
          $T^{*}$ (DFT) & $\delta_{0}^{*}$ & $\delta_{1}^{*}$ & $k_{\textrm{s}}^{*}$\\
          \hline
          0.53 & -0.139 & 0.865 & -0.807\\
          0.61 & -0.147 & 1.082 & -0.717\\
          0.68 & -0.141 & 1.261 & -0.501\\
          0.76 & -0.165 & 1.965 & -0.382\\
          0.83 & -0.164 & 2.886 & -0.285\\
          0.91 & -0.155 & 5.040 & -0.197\\       
          \hline
           $T$ (MD) & $\delta_{0}$ & $\delta_{1}$ & $k_{\textrm{s}}$\\
           \hline
           290 K & -0.59 ($\mathring{\textrm{A}}$) & 9.27 ($\mathring{\textrm{A}}^{2}$) & -1.34 ($k_{\rm{B}}T$)  \\

          \hline
     \end{tabular}
 \end{table}
 
 Using the simulation results of $\delta_{0}$ and $\delta_{1}$ presented in Fig. \ref{fig:fig1}(b) and Table \ref{table:tolman}, we now can derive the analytic governing equation of surface tension from the rigorous Gibbs-Tolman-Koenig-Buff (GTKB) equation \cite{Gibbs1906, Koenig1950, Buff1951, Rowlinson1982},
 \begin{equation}\label{eqn:GTKB}
\frac{\textrm{dln}\gamma}{\textrm{dln}R_{\textrm{s}}} = \frac{\frac{2\delta(R_{\textrm{s}})}{R_{\textrm{s}}}\left[1+\frac{\delta(R_{\textrm{s}})}{R_{\textrm{s}}}+\frac{1}{3}{\left(\frac{\delta(R_{\textrm{s}})}{R_{\textrm{s}}}\right)}^2 \right]} {1+\frac{2\delta(R_{\textrm{s}})}{R_{\textrm{s}}}\left[ 1+\frac{\delta(R_{\textrm{s}})}{R_{\textrm{s}}}+\frac{1}{3}{\left(\frac{\delta(R_{\textrm{s}})}{R_{\textrm{s}}}\right)}^2\right]}.
 \end{equation} 
Here we have used the series expansion using the ansatz $\delta(R_{\textrm{s}}) = \delta_{0} + \delta_{1}/R_{\textrm{s}}$ and thus the resulting equation can be  written as, up to the second order of $1/R_{\textrm{s}}$,
 \begin{equation} 
 \frac{\gamma(R_{\textrm{s}})}{\gamma_{0}} = 1-\frac{2\delta_{0}}{R_{\textrm{s}}}+\frac{3\delta_{0}^{2}-\delta_{1}}{R_{\textrm{s}}^{2}}+O\left({\left(\frac{1}{R_{\textrm{s}}}\right)}^3\right).
 \label{eqn:TtoH}
 \end{equation}
 Equation \eqref{eqn:TtoH} is our  analytic result and is expected to be more exact compared to  Eq. \eqref{eqn:Tolman} since the curvature dependent Tolman length is considered. To demonstrate the validity of Eq. \eqref{eqn:TtoH}, we compare it with the surface tension directly calculated from MD and DFT results without the ansatz (Fig. \ref{fig:fig1}(c)). The surface tension of the TIP4P/2005 water droplet is computed from the difference between the normal and transverse pressure tensors\cite{Ghoufi2011}. Surface tension of Lennard-Jones liquid droplet is calculated from the grand potential energy of the system (refer to Supplementary Information for detailed information). As shown in Fig. \ref{fig:fig1}(c), Eq. \eqref{eqn:TtoH} predicts excellently the surface tension of nanodroplets down to the smallest radius we consider, $R_{\textrm{s}} = 0.45 \ \textrm{nm}$ for water. We estimate  the accuracy of Eq. \eqref{eqn:TtoH}, for example, at $T^{*} = 0.53$ of Lennard-Jones droplet;  Eq. \eqref{eqn:TtoH} shows an error within 0.5 $\%$ until the lower size limit of nanodroplet, while the error jumps over 10 $\%$ without the second order term (Supplementary Fig. 1). The observed Tolman length and surface tension equation with second order correction agrees well with the recent results of Rehner and Gross\cite{Rehner2018}. Our study shows that the Tolman length approach can be applied to the hydrogen-bond liquid with strong directional bond, and also the second-order correction is good for describing surface tension of 1 nm nanodroplet.

 Interestingly, the second-order  term in Eq. \eqref{eqn:TtoH} is closely related to the bending rigidity\cite{Helfrich1973}, which is typically used as an empirical parameter to describe the free energy of the curved membrane. The bending rigidity, $k_{s}$,  can be related to the Tolman length as $k_{\textrm{s}} = \gamma_{0}(3\delta_{0}^{2}-\delta_{1})$\cite{Helfrich1973}.  For water nanodroplets, we find the negative rigidity as shown in Table \ref{table:tolman}, $k_{\textrm{s}} = -1.34 k_{\textrm{B}}T$, which is comparable to  the MD result \cite{Sedlmeier2012}, $k_{\textrm{s}} \approx -3 k_{\textrm{B}}T$, that considered the  interface between hydrophobic solute and water.
Here, the negative sign of bending rigidity is the general feature of the liquid nanodroplet of water, Lennard-Jones liquid, and polymeric liquid \cite{Rehner2018}.

\begin{figure}
\includegraphics[width=8.6cm]{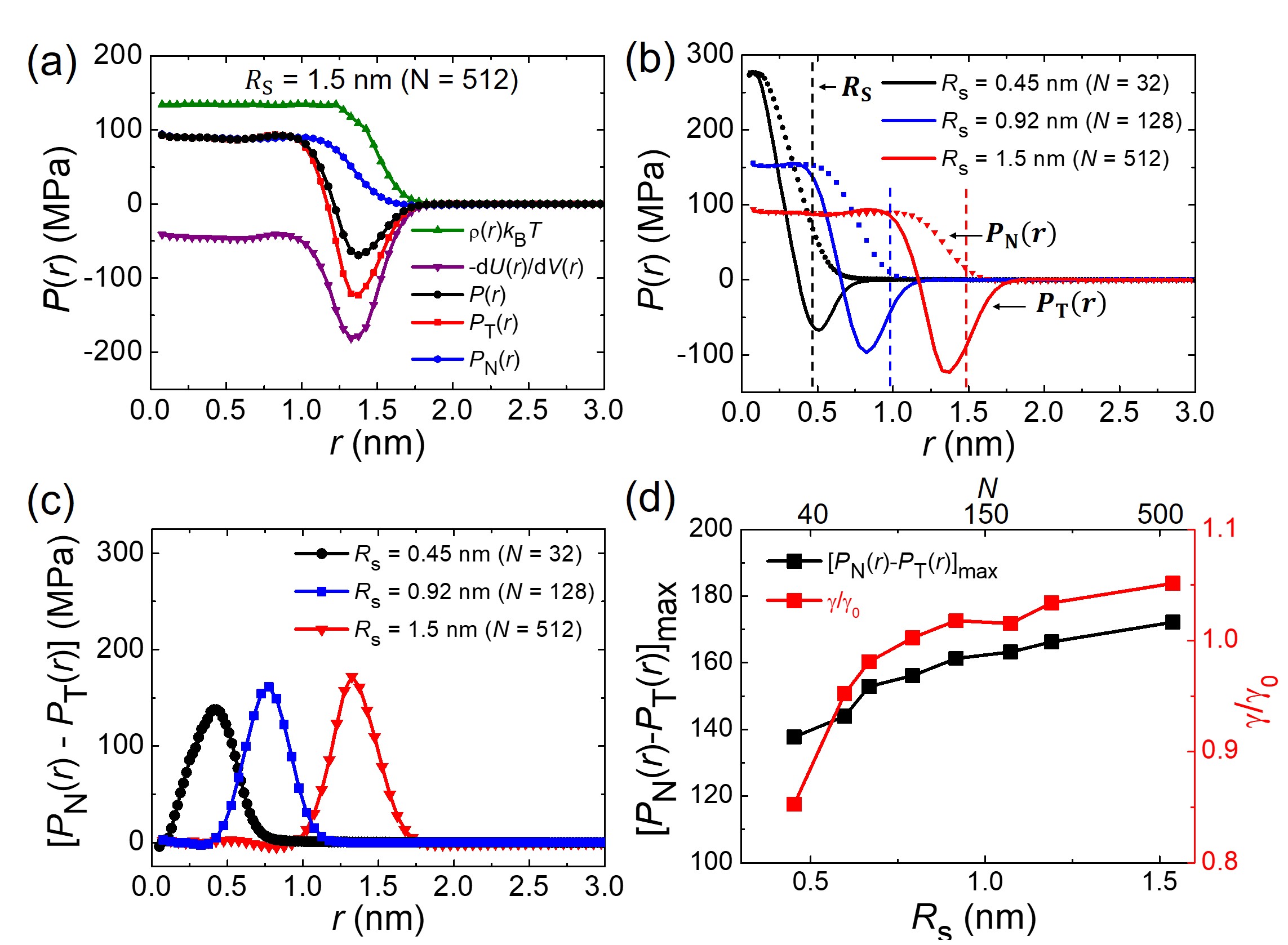}
\centering
\caption{Pressure tensor of nanodroplets. (a) Radial pressure ($P(r)$, black) of $R_{\textrm{s}}$ = 1.5 nm nanodroplet obtained from the sum of kinetic (green) and configurational contribution (purple). $P(r)$ is decomposed into the transverse ($P_{\textrm{T}}(r)$, red) and normal ($P_{\textrm{N}}(r)$, blue) pressure tensor components. (b) Normal and transverse pressure profiles of droplets. The vertical dashed lines indicate the locations of $R_{\textrm{s}}$. (c) Radial profiles of pressure difference ($P_{\textrm{N}}(r) - P_{\textrm{T}}(r)$). (d) The similar behaviors between the maximum values of pressure difference (black) and surface tension (red) shows that the reduced surface tension originates from the pressure imbalance at the interface.}
\label{fig:fig2} 
\end{figure}

Let us now elucidate the physical origin of the smaller surface tension with respect to the planar interface from three perspectives; (i)  pressure tensor, (ii) free energy, and (iii) HB  network.  First, we calculate the radial pressure tensor of molecules, $P(r)$, which comes from the sum of kinetic pressure, $\rho(r)k_{\textrm{B}}T$, obtained from radial density $\rho(r)$ and configurational pressure, $-\left<\Delta U(r) / \Delta V(r) \right>$, obtained from potential energy difference induced by volume perturbation (refer to Supplementary Information for detailed derivation). We  decompose $P(r)$ into the transverse ($P_{\textrm{T}}(r)$) and normal ($P_{\textrm{N}}(r)$) profiles under the hydrostatic equilibrium condition ($\nabla \cdot \vec{P}(r) = 0$) and the boundary condition ($P(r) = 1/3P_{\textrm{N}}(r)+2/3P_{\textrm{T}}(r)$). At the interface of $R_{\textrm{s}}$ = 1.5 nm droplet (Fig. \ref{fig:fig2}(a)), $P_{\textrm{T}}(r)$ deviates from $P_{\textrm{N}}(r)$ and shows a negative minimum, which indicates the origin of surface tension from the imbalanced interaction of the interfacial molecules. Notice that the values of negative minimum (Fig. \ref{fig:fig2}(b)) and pressure imbalance (Fig. \ref{fig:fig2}(c)) decrease as the droplet radius decreases, which shows that the interaction between interfacial molecules weakens accordingly and is responsible for the curvature dependence of surface tension. As expected, a direct comparison between surface tension and pressure imbalance (Fig. \ref{fig:fig2}(d)) evidences the relationship between the two.

\begin{figure}
\includegraphics[width=8.6cm]{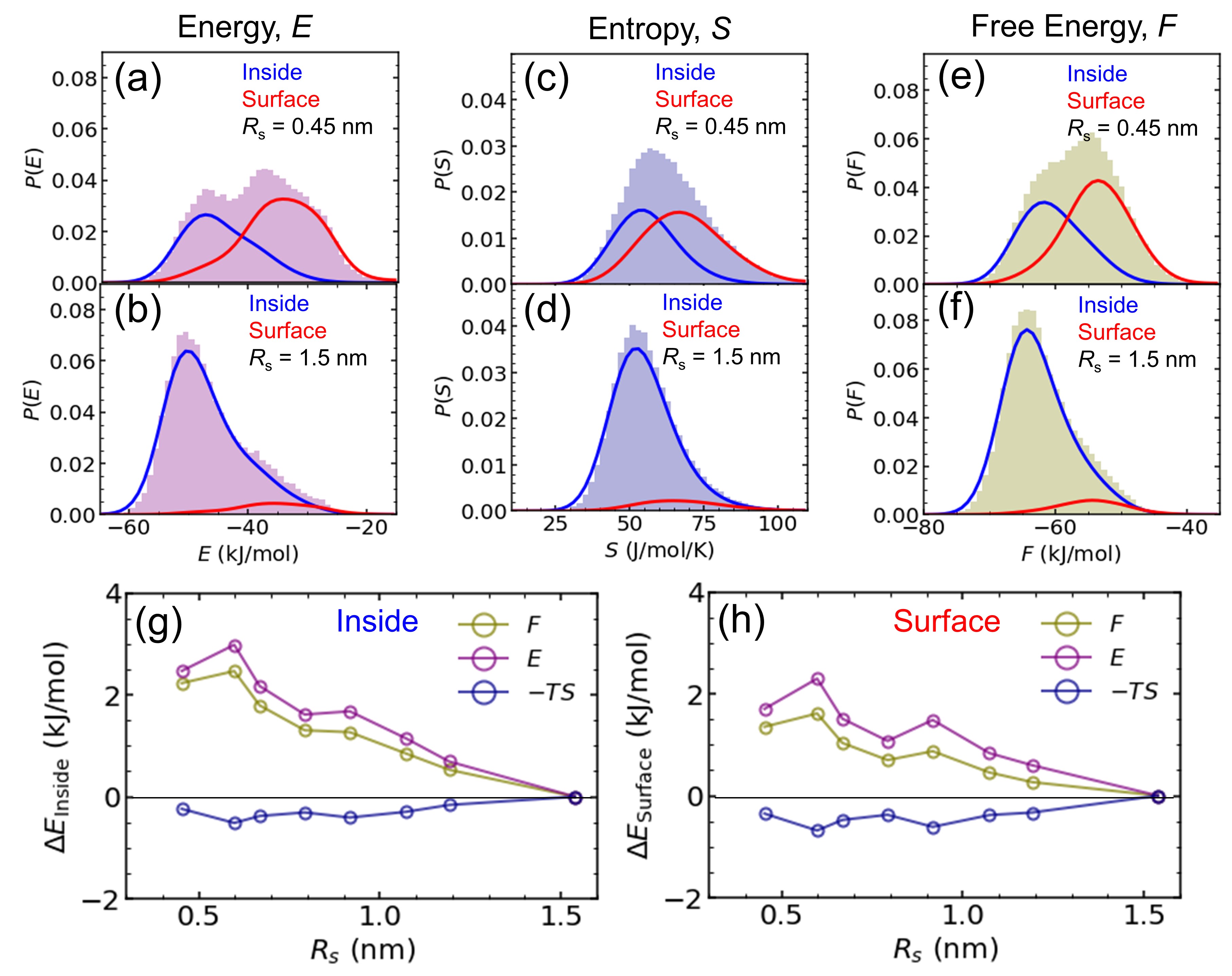}
\centering
\caption{Thermodynamics of nanodroplets. Probability distribution of energy ((a), (b)), entropy ((c), (d)), and Helmholtz free energy ((e), (f)) of water molecules for nanodroplets of  $R_{\textrm{s}}$ = 0.45 nm ($N = 32$) (upper row) and  $R_{\textrm{s}}$ = 1.5 nm ($N = 512$) (lower row). With $R_{\textrm{s}}$, we separate the nanodroplet into the inside region having low energy and low entropy, and the  surface region with high energy and high entropy, which are represented by two solid lines. The average energy, entropy and Helmholtz free energy of molecules of the `inside' region and the `surface' region, presented respectively in (g) and (h), shows that the interfacial energy contributes dominantly to the reduced surface tension.}
\label{fig:fig3} 
\end{figure}

 Second, we investigate the physical mechanism with regard to interfacial thermodynamics at the molecular level. For this purpose, we compute the energy ($E$), entropy ($S$), and Helmholtz free energy ($F$) of nanodroplets (Fig. \ref{fig:fig3}). Entropy is derived from the  integration of the density of states of molecules, which are obtained via the Fourier transform of the velocity autocorrelation function. Formalism is established by the two-phase thermodynamics (2PT) model in calculating the entropy of liquid molecules \cite{Huang2011} (refer to Supplementary Information for detailed information). Helmholtz free energy is calculated by the relation $F = E - TS$. Figures \ref{fig:fig3}(a) - (f) show the probability distributions of $E$, $S$, and $F$  for the droplets of $R_{\textrm{s}}$ = 0.45 nm and 1.5 nm. By dividing two regions across $R_{\textrm{s}}$, each nanodroplet is well separated into the `inside' region having low energy and low entropy, and the `surface' region with high energy and high entropy. Figures \ref{fig:fig3}(g) and (h) compile all the thermodynamic quantities of water molecules in the inside  and  surface region as a function of nanodroplet radius, respectively. For example, the average Helmholtz free energy of the surface molecules for $R_{\textrm{s}}$ = 0.45 nm  is about 2 kJ/mol higher than that of $R_{\textrm{s}}$ = 1.5 nm, which evidences that the change of surface tension is associated with the decrease of intermolecular interaction strength of nanodroplets. Notice that the free energy change is dominated by the energy, which compensates for the entropy change that behaves oppositely.

 \begin{figure}
\includegraphics[width=8.6cm]{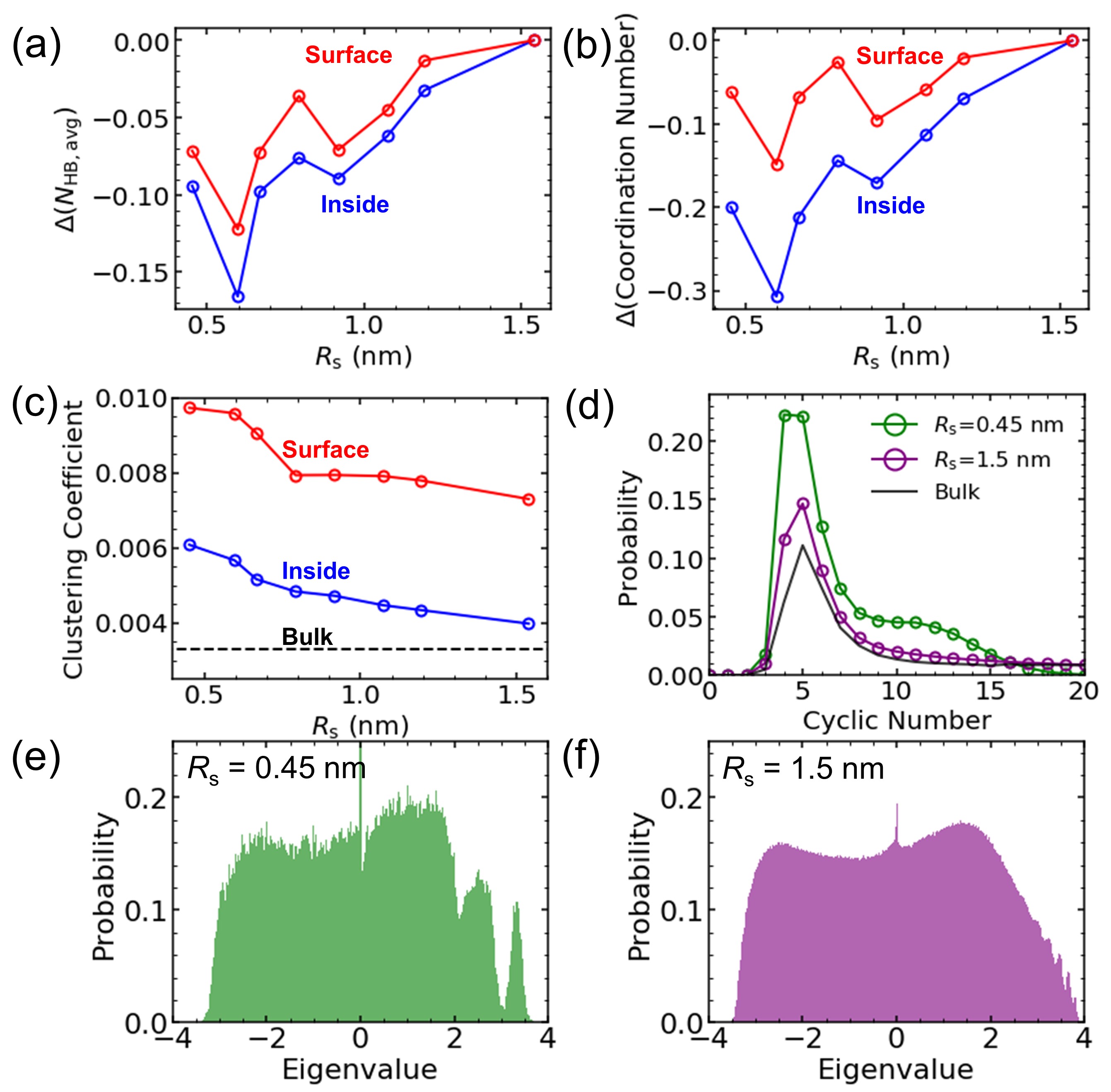}
\centering
\caption{Hydrogen bond (HB) network analysis of nanodroplets. Change of (a) the average number of HB, $\Delta N_{\textrm{HB,avg}}$  and (b) the average coordination number with respect to those of $R_{\textrm{s}}$ = 1.5 nm nanodroplet. (c) Average clustering coefficients of HB networks of nanodroplets in the inside region (blue) and in the surface region (red). Clustering coefficient of the bulk water networks is drawn in black. (d) Probability distributions of the cyclic number of HB networks of $R_{\textrm{s}}$ = 0.45 nm ($N = 32$) and  $R_{\textrm{s}}$ = 1.5 nm ($N = 512$) nanodroplets. Cyclic number distribution of the bulk water network is drawn in black. Probability distributions of eigenvalues of the adjacency matrix associated with the HB networks of (e) $R_{\textrm{s}}$ = 0.45 nm ($N = 32$) and  (f) $R_{\textrm{s}}$ = 1.5 nm ($N = 512$) nanodroplets.}\label{fig:fig4}
\end{figure}

Lastly, we address the molecular description in terms of the relation between the configurational energy change of nanodroplets and the change of HB network characteristics (refer to Supplementary Information for detailed information). Two water molecules are considered to be connected via HB when (i) the distance between two oxygen atoms is smaller than 0.35 nm and (ii) the angle between hydrogen and donor-acceptor oxygen atoms is smaller than $30^{\circ}$ \cite{Kumar2007}. We represent the HB network of nanodroplets by the adjacency matrix $\bf{A}$, which is a square symmetric matrix with elements of 1 if the molecular pair $(i, j)$ has an HB, and 0 if otherwise. Figure \ref{fig:fig4}(a) shows the change of average number of HB, $N_{\textrm{HB,avg}}$, for the inside as well as the surface molecules in comparison to  $N_{\textrm{HB,avg}}$ of $R_{\textrm{s}}$ = 1.5 nm nanodroplet. We find that the smaller nanodroplets have relatively smaller $N_{\textrm{HB,avg}}$, which is already expected by Figs. \ref{fig:fig3}(g) and (h). The disruption of the HB network shown in Fig. \ref{fig:fig4}(a) can be explained by the decreasing coordination number (Fig. \ref{fig:fig4}(b)). Notice that these results are independent of the choice of radius, which is still observed even if $R_{\textrm{e}}$ is used instead of $R_{\textrm{s}}$ (Supplementary Fig. 2). The overall change of the HB network characteristics can be observed by the clustering coefficient and cyclic number distribution of the  network presented in Figs. \ref{fig:fig4}(c) and (d), respectively. The clustering coefficient of the HB network is defined as the probability that the two hydrogen-bonded neighboring molecules are connected via the HB and form a triangular network. The cyclic number is defined as the number of edges connected in a closed HB loop in a network.  As shown, the nanodroplet network has a larger average clustering coefficient and a smaller average cyclic number than the bulk water network, which demonstrates that the nanodroplets are more structured than the bulk water, and their structure is similar to that of the high-density water or Ice-VII \cite{Shin2019}. Moreover, from the eigenvalue spectra of $\bf{A}$ (Figs. \ref{fig:fig4}(e) and (f)), one can observe that the change of network structure is concentrated in the local network topology with large eigenvalue of $\bf{A}$ (or large number of HB), which shows that the disruptions of HB networks are focused on the four or five HBs. Because of the large energy content of water HB, disruption of the HB network directly leads to the increase of free energy and the resulting decrease of surface tension.

As a conclusion, we emphasize that the above three perspectives provide the unified understanding of the decreasing surface tension in terms of the decrease of interaction strength (Fig. \ref{fig:fig2}), an increase of energy (Fig. \ref{fig:fig3}) and the decrease of HB number as well as the change of HB network structure  (Fig. \ref{fig:fig4}) in the molecular-scale nanodroplets. In particular, the large HB energy of water (i) compensates the entropy change and (ii) dominates the change of Helmholtz free energy and the change of pressure imbalance.

In summary, we have demonstrated  the curvature dependence of surface tension  from the perspectives of interfacial thermodynamics  using the complementary methods of MD and DFT simulations. We show that the surface tension  is a decreasing function of  the inverse radius of nanodroplets with the positive Tolman length. Statistical analysis of the Helmholtz free energy of individual water molecules shows that the change of surface tension is governed by the surface molecules  having higher free energy than that of the planar interface, which is closely related to the disruption of the HB network at the interface. Our results may trigger further theoretical development for an accurate understanding of the nucleation of nanoclusters at the molecular level. Moreover, the thermodynamic approach presented here can be used to study the chemically heterogenous interface where the local energy varies at the nanoscale.

\section*{Acknowledgements}
The authors are grateful to Dr. Fang Gu for introduction of the DFT formalism of the Lennard-Jones liquid. This work was supported by the National Research Foundation of Korea (NRF) grant funded by the Korea Government (MSIP) (No. 2016R1A3B1908660).



\begin{thebibliography}{tbhp}
\bibitem{Baker1997} M. Baker, {Cloud microphysics and climate,} Science \textbf{276}, 1072-1078 (1997).

\bibitem{Kulmala2004} M. Kulmala, H. Vehkam{\"a}ki, T. Pet{\"a}j{\"a}, M. Dal Maso, A. Lauri, V. -M. Kerminen, W. Birmili, and P. H. McMurry, {Formation and growth rates of ultrafine atmospheric particles: a review of observations,} J. Aerosol Sci. \textbf{35}, 143-176 (2004).

\bibitem{Kalikmanov2005} V. Kalikmanov, {Nucleation Theory} (Spinger Netherlands, 2013).

\bibitem{Bruot2016} N. Bruot, and F. Caupin, {Curvature dependence of the liquid-vapor surface tension beyond the Tolman approximation,} Phys. Rev. Lett. \textbf{116}, 056102 (2016).


\bibitem{Oxtoby1988} D. W. Oxtoby, {Nonclassical nucleation theory for the gas-liquid transition,} J. Chem. Phys. \textbf{89}, 7521 (1988).

\bibitem{Holten2005} V. Holten, D. G. Labetski, and M. E. H. van Dongen, {Homogeneous nucleation of water between 200 and 240 K: New wave tube data and estimation of the Tolman length,} J. Chem. Phys. \textbf{123}, 104505 (2005).

\bibitem{Sedlmeier2012} F. Sedlmeier, and R. R. Netz, {The spontaneous curvature of the water-hydrophobe interface,} J. Chem. Phys. \textbf{137}, 135102 (2012).

\bibitem{Kim2018} S. Kim, D. Kim, J. Kim, S. An, and W. Jhe, {Direct evidence for curvature-dependent surface tension in capillary condensation: Kelvin equation at molecular scale,} Phys. Rev. X \textbf{8}, 041046 (2018).

\bibitem{Kwon2018} S. Kwon, B. Kim, S. An, W. Lee, H. Kwak, and W. Jhe, {Adhesive force measurement of steady-state water nano-meniscus: Effective surface tension at nanoscale,} Sci. Rep \textbf{8}, 8462 (2018).

\bibitem{Factorovich2014} M. H. Factorovich, V. Molinero, and D. A. Scherlis, {Vapor pressure of water nanodroplets,} {J. Am. Chem. Soc.} 136, 4508-4514 (2014).

\bibitem{Sampayo2010} J. G. Sampayo, A. Malijevsky, E. A. Muller, E. de Miguel, and G. Jackson, {Communications: Evidence for the role of fluctuations in the thermodynamics of nanoscale drops and the implications in computations of the surface tension,} J. Chem. Phys. \textbf{132}, 141101 (2010).

\bibitem{Koga1998} K. Koga, X. C. Zeng, and A. K. Shchekin, {Validity of Tolman's equation: How large should a droplet be?} J. Chem. Phys. \textbf{109}, 4063 (1998).

\bibitem{Malijevsky2012} A. Malijevsky, and G. Jackson, {A perspective on the interfacial properties of nanoscopic liquid drops,} J. Phys.: Condens. Matter \textbf{24}, 464121 (2012).

\bibitem{Rehner2018} P. Rehner and J. Gross, {Surface tension of dropelts and Tolman lengths of real substances and mixtures from density functional theory,} J. Chem. Phys. \textbf{148}, 164703 (2018).

\bibitem{Rowlinson1982} J. S. Rowlinson, and B. Widom, {Molecular theory of capillarity.} (Clarendon Press, 1982).

\bibitem{Wilhelmsen2015} O. Wilhelmsen, D. Bedeaux, and D. Reguera, {Communication: Tolman length and rigidity constants of water and their role in nucleation,} J. Chem. Phys. \textbf{142}, 171103 (2015).

\bibitem{Azouzi2013} M. E. M. Azouzi, C. Ramboz, J. F. Lenain, and F. Caupin, {A coherent picture of water at extreme negative pressure,} Nat. Phys. \textbf{9}, 38-41 (2013).

\bibitem{Kim2018-2} Q. Kim, D. Shin, J. Park, D. A. Weitz, and W. Jhe, {Initial growth dynamics of 10 nm nanobubbles in the graphene liquid cell,} Appl. Nanosci. https://doi.org/10.1007/s13204-018-0925-3 (2018).

\bibitem{Joswiak2013}  M. N. Joswiak, N. Duff, M. F. Doherty, and B. Peters, {Size-dependent surface free energy and Tolman-corrected droplet nucleation of TIP4P/2005 water,} J. Phys. Chem. Lett. \textbf{4}, 4267-4272 (2013).

\bibitem{Tolman1949} R. C. Tolman, {The effect of droplet size on surface tension,} J. Chem. Phys. \textbf{17}, 333 (1949).

\bibitem{Vega2006} C. Vega, J. L. F. Abascal, and I. Nezbeda, {Vapor-liquid equilibria from the triple point up to the critical point for the new generation of TIP4P-like models: TIP4P/Ew, TIP4P/2005, and TIP4P/ice,} J. Chem. Phys. \textbf{124}, 074507 (2006).

\bibitem{Block2010} B. J. Block, S. K. Das, M. Oettel, P. Virnau, and K. Binder, {Curvature dependence of surface free energy of liquid drops and bubbles: A simulation study,} J. Chem. Phys. \textbf{133}, 154702 (2010).

\bibitem{Gibbs1906} J. W. Gibbs, H. A. Bumstead, and R. G. van Name, {Scientific Papers of J. Willard Gibbs: Thermodynamics} (Longmans and Green, New York).

\bibitem{Koenig1950} F. O. Koenig, {On the thermodynamic relation between surface tension and curvature,} J. Chem. Phys. \textbf{18}, 449 (1950).

\bibitem{Buff1951} F. P. Buff, {The spherical interface. I. Thermodynamics,} J. Chem. Phys. \textbf{19}, 1591 (1951).

\bibitem{Ghoufi2011} A. Ghoufi, and P. Malfreyt, {Local pressure components and surface tension of spherical interfaces. Thermodynamic versus mechanical definitions. I. A mesoscale modeling of droplets,} J. Chem. Phys. \textbf{135}, 104105 (2011).

\bibitem{Helfrich1973} W. Helfrich, {Elastic properties of lipid bilayers: theory and possible experiments,} Z. Naturforsch. C \textbf{28}, (1973).

\bibitem{Huang2011} S. N. Huang, T. A. Pascal, W. A. Goddard, P. K. Maiti, and S. T. Lin, {Absolute entropy and energy of carbon dioxide using the two-phase thermodynamic model,} J. Chem. Theory Comput. \textbf{115}, 14190-14195 (2011).

\bibitem{Kumar2007} R. Kumar, J. R. Schmidt, J. L. Skinner, Hydrogen bonding definitions and dynamics in liquid water. \textit{J. Chem. Phys.} \textbf{126}, 204107 (2007).

\bibitem{Shin2019} D. Shin, J. Hwang, and W. Jhe, {Ice-VII-like molecular structure of ambient water nanomeniscus,} Nat. Comm. \textbf{10}, 286 (2019).

\end{thebibliography}
\end{document}



\title{Supplementary Information : Interfacial thermodynamics of spherical nanodroplets: Molecular understanding of surface tension via hydrogen bond network}\author{QHwan Kim}
\author{Wonho Jhe}%
 \email{whjhe@snu.ac.kr}
\affiliation{%
 Center for 0D Nanofluidics, Department of Physics and Astronomy, Institute of Applied Physics,  Seoul National University, Gwanak-gu, Seoul 08826, Republic of Korea.
}%

\maketitle


\section*{Molecular Dynamics Simulation and Surface tension}

We used the nanodroplets of  $N$ = 32 to $N$ = 512 TIP4P/2005 water molecules \cite{Abascal2005}. The O-H bonds length and the H-O-H angle of  water molecules are constrained using the SHAKE algorithm \cite{Ryckaert1977} with a relative tolerance of $10^{-5}$. We used the canonical ensemble, where the total number of molecules $N$, the volume of the system $V$, and the temperature of the system $T$ are conserved. Temperature was controlled using the Nose-Hoover thermostat with a time constant of 1 ps. The droplets were initially located in the center of simulation box under periodic boundary conditions in every direction. We chose each length of cubic simulation box as $L$ = 20 nm, which is large enough to avoid any long-range electrostatic interactions between the original and its periodic images. We employed the cut-off scheme for computing both the Lennard-Jones and the Coulomb interactions without any further correction. The cut-off length was set as 9.5 nm that is about half of the simulation cell length. The equations of motion were integrated using the leap-frog algorithm at a time step of 1 fs. The position of all the molecules were stored every 0.1 ps during 100 ns simulation and we collected $10^{6}$ snapshots for analysis. All simulations were carried out with GROMACS 5.1.4   \cite{Pronk2013}.

 To obtain the surface tension of nanodroplet, we first calculate the pressure tensor of system from the volume perturbation method (Detailed method is discussed in \textit{Pressure Profiles} section). Using the transverse and normal pressure tensor, we obtain the surface tension $\gamma$ from the Lovett's expression as \cite{Lovett1996},
   \begin{equation}
   \gamma = \int_{0}^{\infty} dr \left[ P_{\textrm{N}}(r) - P_{\textrm{T}}(r) \right].
   \end{equation}
 Also the radius of the  surface of tension $R_{\textrm{s}}$ and the equimolar radius $R_{\textrm{e}}$ are determined, respectively, by
 \begin{equation}
R_{\textrm{s}} = \frac{2\gamma}{P_{\textrm{l}}},
 \end{equation} 
   \begin{equation}
 {R_{\textrm{e}}}^3 = \frac{3}{\Delta\rho}\int_{0}^{\infty}\textrm{d}r r^{2}\left[\rho\left(r\right)-\rho_{\textrm{v}}\right],
 \label{eqn:re}
 \end{equation}
 where $P_{\textrm{l}}$ is the pressure at the center of droplets and $\rho_{\textrm{v}}$ is the density of the bulk vapor.

 To calculate the energy and entropy of water molecules using the two phase-themodynamics (2PT) model, we carried out 1.6 ns additional simulation where the position and velocity of all molecules are saved at every 4 fs. We then split the total trajectory into $10^{3}$ sub-trajectories with 1.6 ps length and obtain the average energy and density of states of every single water molecules along  each sub-trajectory.

\section*{Density Functional Theory and surface tension}
 In the density functional theory (DFT) formalism, the grand potential $\Omega\left[\rho\left(\textbf{r}\right)\right]$ corresponding to the density profile $\rho\left(\textbf{r}\right)$ is expressed as, 
\begin{equation}\label{eqn:grand_potential}
\Omega\left[ \rho\left( \textbf{r} \right) \right] = F\left[ \rho\left( \textbf{r} \right)\right] - \mu\int \rm{d\bf{r}}\rho\left( \textbf{r} \right),
\end{equation}
where $F$ is the Helmholtz free energy and $\mu$ is the chemical potential of the system. The Helmholtz free energy, which is divided into the hard sphere reference energy and the dispersive energy term,
\begin{equation}
 F\left[ \rho\left( \textbf{r} \right)\right] = F_{\textrm{hs}}\left[ \rho\left( \textbf{r} \right)\right] + \frac{1}{2}\int \rm{d\textbf{r}}\int \rm{d\textbf{r}}^{\prime} \rho\left( \textbf{r} \right)\rho\left( \textbf{r} ^\prime \right) \textit{U}_{\textrm{disp}}\left( \left| \textbf{r}-\textbf{r}^{\prime} \right| \right),
\end{equation}
 We used the mean field approach to calculate the dispersive energy term, which can be modeled by the Lennard-Jones potential \cite{Weeks1971},
\begin{equation}
U_{\textrm{disp}}\left( r \right)=
\begin{cases} 
-\epsilon & r \leq r_{\textrm{min}}, \\
4\epsilon\left[ {\left(\frac{\sigma}{r} \right)}^{12} - {\left(\frac{\sigma}{r} \right)}^{6} \right] & r_{\textrm{min}} < r < r_{\textrm{c}}, \\
0 & r > r_{\textrm{c}},
 \end{cases}
\end{equation}
 where $\sigma$ is the diameter of the liquid molecule and $r_{\textrm{min}} = 2^{1/6}\sigma$, $r_{\textrm{c}} = 5\sigma$. To calculate the free energy of a hard sphere contribution in Eq. 7,  we used the Carnahan-Starling form \cite{Carnahan1970},
 \begin{equation}
 F_{\textrm{hs}}\left(\rho\right) = k_{\textrm{B}}T\rho\textrm{ln}\left(\rho\right) + k_{\textrm{B}}T\rho\frac{\left(4\eta-3\eta^{2}\right)}{{\left(1-\eta\right)}^{2}},
 \end{equation}
 where $\eta$ is defined as  $\eta = \left(\pi/6\right)\rho \sigma^{3}$. In general, the density profile of the equilibrium system can be obtained by finding the minimum value of the grand potential for which  the functional derivative of  Eq. \ref{eqn:grand_potential} becomes zero, so that 
\begin{equation}
\mu = \frac{\delta F_{\textrm{hs}}\left[\rho\left(\textbf{r}\right)\right]}{\delta \rho\left(\textbf{r}\right)} + \int\rm{d\textbf{r}}^{\prime} \rho\left(\textbf{r}^{\prime}\right) \textit{U}_{\textrm{disp}}\left( \left| \textbf{r}-\textbf{r}^{\prime} \right| \right).
\label{eqn:DFT}
\end{equation}

The initial density profile of nanodroplet was located in the center of closed canonical system and was updated via Picard iteration until the system reaches the minimum free energy state \cite{Malijevsky2012}.

We obtain the surface tension $\gamma(R_{\textrm{s}})$ from the grand potential and the corresponding pressure of the  system:
 
\begin{equation}
\Delta\Omega=-\frac{4 \pi\Delta P}{3}R_{\textrm{s}}^{3}+4\pi R_{\textrm{s}}^{2} \gamma\left(R_{\textrm{s}}\right),
\label{eqn:tension}
\end{equation}
 where $\Delta\Omega = \Omega+P_{\textrm{v}}V$ and $\Delta P = P_{\textrm{l}}-P_{\textrm{v}}$. The radius of the surface of tension $R_{\textrm{s}}$ is then obtained from its definition ${{\textrm{d}\gamma}/{\textrm{d}R}}|_{R=R_{\textrm{s}}}=0$ as,
 \begin{equation}
 {R_{\textrm{s}}}^3 = {\frac{3\Delta\Omega}{2\pi\Delta P}}.
 \label{eqn:rs}
 \end{equation}
Notice that the equimolar radius $R_{\textrm{e}}$ can be obtained by the same equation as in the MD simulation (Eq. \ref{eqn:re}). 

 All the DFT calculations were performed using the home-made code implemented by the C language.

\section*{Pressure Profiles}
To calculate the normal ($P_{\textrm{N}}(r)$) and tangential ($P_{\textrm{T}}(r)$) components of the radial pressure tensor ($P(r)$), we used the volume perturbation method \cite{Ghoufi2011}. 
 We first compute $P(r)$ by the relation,
  \begin{equation}
  \label{eqn:local_pressure2}
  P(r) = \rho(r)k_{\textrm{B}}T - \left < \frac{\partial U(r)}{\partial V(r)}\right >_{NVT},
  \end{equation}
  where $\rho(r)$ is the radial density profile, $U(r)$ is the radial potential energy profile and $V(r)$ is the radial profile of the spherical shell.  Equation \ref{eqn:local_pressure2} shows that the local pressure is the summation of the local-density energy and the potential energy change with respect to change of volume \cite{Gloor2005}. To change the volume of the spherical nanodroplet, compressing and expanding from a reference state (0) to a perturbed state (1) are used according to the isotropic transformation of the center of mass: $\textbf{r}_{\textrm{CM}} \rightarrow (1 + \xi)\textbf{r}_{\textrm{CM}}$. Using this approach, Eq. \ref{eqn:local_pressure2} can be rewritten as, 
  \begin{equation}
  \label{eqn:local_pressure3}
  P(r) = \rho(r)k_{\textrm{B}}T -  \lim_{\xi \rightarrow 0} \left< \frac{ U^{1}(r) - U^{0}(r)}{V^{1}(r) - V^{0}(r)} \right>_{NVT},
  \end{equation}
 where the rescaled volume of the spherical shell is $V^{1}(r) = (1+\xi)^{3}V^{0}(r)$. 
 The perturbation of the spherical system is defined as volume compression ($\xi < 0$) or volume expansion ($\xi > 0$) from a reference state (0) to a perturbed state (1) according to the isotropic transformation of the center of mass; $\textbf{r}_{\textrm{CM}} \rightarrow (1 + \xi)\textbf{r}_{\textrm{CM}}$. The calculation of local elements of the pressure tensor is carried out by averaging the values of two perturbed states. The ensemble average is then performed over the configurations of the reference system 0, for which we use $\xi = 10^{-5}$. 
  
  To decompose $P(r)$ into $P_{\textrm{N}}(r)$ and $P_{\textrm{T}}(r)$, we used the mechanical equilibrium condition of spherical geometry \cite{Thompson1984}, 
 \begin{equation}
 \label{eqn:equil1}
 P(r) = \frac{1}{3}P_{\textrm{N}}(r) + \frac{2}{3}P_{\textrm{T}}(r),
 \end{equation}
 \begin{equation}
 \label{eqn:equil2}
 P_{\textrm{T}}(r) = P_{\textrm{N}}(r) + \frac{r}{2}\frac{\textrm{d}P_{\textrm{N}}(r)}{\textrm{d}r}.
 \end{equation}
 Combining Eq. \ref{eqn:equil1} and Eq. \ref{eqn:equil2}, we then obtain  $P_{\textrm{N}}(r)$ as,
 \begin{equation}
 P_{\textrm{N}}(r) = \frac{1}{r^{3}}\int_{0}^{r}3r'^{2}P(r')\textrm{d}r',
 \end{equation}
 for which $P_{\textrm{T}}(r)$ can be calculated from Eq. \ref{eqn:equil2}.

 \section*{Two-Phase Thermodynamics Model}
 We compute the entropy of water molecules by integrating the density of states (DoS) using the two-phase thermodynamics (2PT) model, which was first introduced by  Lin \textit{et al.} \cite{Lin2010}. The 2PT model considers the liquid system as the sum of diffusive gas-like component and solid-like harmonic spring component. We first calculate the DoS from the Fourier transformation of the velocity autocorrelation function as,
 \begin{equation}
 g(\nu) = \lim_{\tau \rightarrow \infty} \int_{-\tau}^{\tau} C(t) \textrm{exp}(-i2\pi \nu t), 
 \end{equation}
 where $C(t)$ is the autocorrelation function, defined by,
 \begin{equation}
 C(t) = \sum_{j} \lim_{\tau \rightarrow \infty} \frac{1}{2\tau} \int_{-\tau}^{\tau} \textrm{d}t' v_{j}(t+t') v_{j}(t).
 \end{equation}
 When the system is solid-like, harmonic oscillation dominates the mode of dynamics while diffusion is neglected. However, the DoS of liquid has a non-zero value of diffusional components DoS(0), which leads to incorrect entropy evaluation when one considers the system of liquid molecules only as the sum of harmonic oscillators. To avoid this problem, DoS is decomposed into  diffusive gas-like and harmonic solid-like components,
 \begin{equation}
 g(\nu) = g^{g}(\nu) + g^{s}(\nu),
 \end{equation}
 where $g^{g}(\nu)$ is the diffusive component and $g^{s}(\nu)$ is the harmonic component. Here,  $g^{g}(\nu)$ is expressed as,
 \begin{equation}
 g^{g}(\nu) = \frac{g^{g}(0)}{1 + \left[ \frac{\pi g^{g}(0)\nu}{6fN} \right]^{2}},
 \end{equation}
 where $f$ denotes the total degrees of freedom of the gas-like components, $g^{g}(0)$ is the DoS at zero frequency, and $N$ is the number of molecules.
 For polyatomic molecules (e.g., water), the dynamics is decomposed into the translational, rotational and vibrational components given by,
 \begin{equation}
 g^{i}(\nu) =  g^{i}_{\textrm{trn}}(\nu) +  g^{i}_{\textrm{rot}}(\nu) +  g^{i}_{\textrm{vib}}(\nu).
 \end{equation}
 Notice that  $g^{i}_{\textrm{trn}}(\nu)$ is calculated from the translational velocities of the center of mass of molecules, $g^{i}_{\textrm{rot}}(\nu)$  from the angular velocity of molecules, and $g^{i}_{\textrm{vib}}(\nu)$  from the intramolecular vibrations. Here, $g^{i}_{\textrm{vib}}(\nu) = 0$ because the rigid TIP4P/2005 water shows no intramolecular vibration.
 
  As a result, the entropy $S$ can be obtained by the sum of four contributions,
 \begin{equation}
 S = S^{g}_{\textrm{trn}} + S^{g}_{\textrm{rot}} + S^{s}_{\textrm{trn}} + S^{s}_{\textrm{rot}},
 \end{equation}
 and each component is expressed by  integration of DoS with the weighting function,
 \begin{equation}
 S^{i}_{j} = \int_{0}^{\infty} \textrm{d}\nu g^{i}_{j}(\nu) W^{i}_{j}(\nu),
 \end{equation}
 where $i = \{ s, g \}$ and $j = \{ \textrm{trn}, \textrm{rot}\}$. Details of the weight functions can be found in work of Lin \textit{et al.} \cite{Lin2010}

 \section*{Graphical construction of hydrogen bond network}
We constructed a graph network based on the HB connection between water molecules. The HB graph, $G_{\textrm{HB}} = \left( \bf{V}, \bf{E} \right)$, is composed of the vertices, $\textbf{V} = \{1, 2, ..., N\}$ (or the set of the molecular index) and the edges, $\textbf{E} = \{(i, j)\}$ (or the set of the HB-connected pair). Since $G_{\textrm{HB}}$ is the undirected simple graph, we can use the symmetric adjacency matrix $\textbf{A}$ to represent the graphical information such that,
\begin{equation}
A_{ij} =
\begin{cases} 
0 & i = j, \\
1 & i \neq j, \rm{if \ HB \ exists}, \\
0 & \rm{otherwise}.
\end{cases}
\end{equation}
Here, two water molecules are considered to be connected via HB when (i) the distance between two oxygen atoms are smaller than 0.35 nm and (ii) the angle between hydrogen and donor-acceptor oxygen atoms is smaller than $30^{\circ}$ \cite{Kumar2007}. $\textbf{A}$ contains the informations of $\textbf{V}$, $\textbf{E}$ and is a format that facilitates graphical analysis. For example, the number of HB of the $i$-th molecule, $N_{\textrm{HB}}(i)$, can be simply obtained by summation of the $i$-th row elements of $\bf{A}$.

The clustering coefficient of the HB network graph is defined as,
\begin{equation}
\textrm{Cluster. Coeff.}(i) = \sum_{j \in A_{ij}=1, k \in A_{ik}=1}\frac{A_{jk}}{N_{\textrm{HB}}(i) \left(N_{\textrm{HB}}(i)-1\right)},
\end{equation}
where the clustering coefficient of molecule $i$ represents the probability that the two molecules $j$ and $k$, which are hydrogen-bonded with $i$, are connected via the hydrogen bond and form the triangular network. Notice that the smaller this value is, the closer it is to the random network and therefore the structured network has a higher value.
Here, the cyclic number is defined as the number of edges connected in a closed HB loop in a network.

The topology of graph can be represented by the eigenvalues $\lambda_{i}$ and eigenvectors $\textbf{l}_{i}$ of the square matrix $\bf{A}$,
\begin{equation}
\textbf{A}\textbf{l}_{i} = \lambda_{i}\textbf{l}_{i} .
\end{equation}
By comparing the eigenvalues of two different graphs, we can analyze how similar the two graphs are and where they differ. Large eigenvalue corresponds to the node with the high $N_{\textrm{HB}}$.


 \section*{Accuracy of Equation of Surface Tension}
 In the main text, we derive the equation of  surface tension, $\gamma(R_{\textrm{s}}) = \gamma_{0} - 2\gamma_{0}\delta_{0}/R_{\textrm{s}} + k_{\textrm{s}}/R_{\textrm{s}}^{2}$. Here let us  estimate quantitatively the accuracy of the equation with respect to the DFT results. Figure \ref{fig:accuracy} shows the accuracy and error between the DFT results at $T^{*} = 0.53$ and predictions obtained by the equation up to three different orders; $\gamma(R_{\textrm{s}}) = \gamma_{0}$, $\gamma(R_{\textrm{s}}) = \gamma_{0} - 2\gamma_{0}\delta_{0}/R_{\textrm{s}}$ and $\gamma(R_{\textrm{s}}) = \gamma_{0} - 2\gamma_{0}\delta_{0}/R_{\textrm{s}} + k_{\textrm{s}}/R_{\textrm{s}}^{2}$. To quantify the error, we use the metric $\rm{Err} = (\gamma_{\rm{DFT}} - \gamma_{\rm{Eq.}})/\gamma_{\rm{DFT}}$, where $\gamma_{\rm{DFT}}$ is derived by DFT and $\gamma_{\rm{Eq}}$ is the prediction from the equations. Figure \ref{fig:accuracy} (b) shows that the equation $\gamma(R_{\textrm{s}}) = \gamma_{0} - 2\gamma_{0}\delta_{0}/R_{\textrm{s}} + k_{s}/R_{\textrm{s}}^{2}$ has a remarkable accuracy with Err $<$ 0.5 $\%$, in comparison with the other two lower-order equations. It clearly demonstrates the importance of second order term for the accurate prediction of surface tension.
